\documentclass[12pt]{article}
\usepackage{setspace}

\usepackage{amsmath, amssymb}
\usepackage[usedvipsnames]{xcolor}
\usepackage{booktabs}
\usepackage{mathrsfs}
\usepackage{multirow}
\usepackage{algorithmicx, algpseudocode, algorithm}
\usepackage{graphicx}
\usepackage{bm}
\usepackage[font=singlespacing]{caption}
\usepackage[margin=1in]{geometry}

\newcommand{\bt}[1]{\begin{center}\begin{tabular}{#1}\hline}
\newcommand{\et}{\\\hline\end{tabular}\end{center}}

\newcommand{\Gam}{\operatorname{Gamma}}

\newcommand{\bfx}{{\bf x}}

\newcommand{\Var}{{\rm Var}}

\newcommand{\Poisson}{\operatorname{Poisson}}
\newcommand{\Uniform}{\operatorname{Uniform}}
\newcommand{\RMST}{\operatorname{RMST}}

\title{Semiparametric Analysis of Clustered Interval-Censored Survival Data using Soft Bayesian Additive Regression Trees (SBART)}

\author{Piyali Basak$^{1,*}$, Antonio Linero$^{2}$, Debajyoti Sinha$^{1}$, Stuart Lipsitz$^{3}$ \\
	{\small $^{1}$\it Florida State University;
	$^2$\it University of Texas at Austin;
	$^{3}$\it Brigham \& Women's Hospital
	}\\
	{\small \it *email: \href{mailto:pb15d@my.fsu.edu}{pb15d@my.fsu.edu}}}

\usepackage[colorlinks = true, citecolor = blue]{hyperref}

\begin{document}

\maketitle

\begin{abstract}
Popular parametric and semiparametric hazards regression models for clustered survival data are inappropriate and inadequate when the unknown effects of different covariates and clustering are complex. This calls for a flexible modeling framework to yield efficient survival prediction. Moreover, for some survival studies involving time to occurrence of some asymptomatic events, survival times are typically interval censored between consecutive clinical inspections.  In this article, we propose a robust semiparametric model for clustered interval-censored survival data under a paradigm of Bayesian ensemble learning, called Soft Bayesian Additive Regression Trees or SBART (Linero and Yang, 2018), which combines multiple  sparse (soft) decision trees to attain excellent predictive accuracy.  We develop a novel  semiparametric hazards regression model by modeling the hazard function as a product of a parametric baseline hazard function and a nonparametric component that uses SBART to incorporate clustering, unknown functional forms of the main effects, and interaction effects of various covariates. In addition to being applicable for left-censored, right-censored, and interval-censored survival data, our methodology is implemented using a data augmentation scheme which allows for existing Bayesian backfitting algorithms to be used. We illustrate the practical implementation and advantages of our method via simulation studies and an analysis of a prostate cancer surgery study where dependence on the experience and skill level of the physicians leads to clustering of  survival times. We conclude by discussing our method's applicability in studies involving high dimensional data with complex underlying associations.
\end{abstract}

\vspace{1em}

\doublespacing

\section{Introduction}

Interval censored survival data arise frequently in asymptomatic diseases that have no immediate outward symptoms (Sun, 2006) and the event of interest, such as device failure or relapse of a disease after initial treatment, is known to occur only between two consecutive inspection times. For instance, recurrence of biomarkers are often detected either during scheduled clinic visits or via some invasive procedures/diagnostic tests causing the survival times to such events to be interval-censored within consecutive inspection times with opposite diagnosis/test results. Our motivation behind this article is to compare two competing surgical techniques, Robotic versus non-robotic Radical Retropubic Prostatectomy (RRP), for prostate removal in terms of the patients' time to recurrence (survival time of interest) of Prostate Specific Antigen (PSA) after the surgery (Barbash and Glied, 2010). 
The PSA level in the blood after prostate removal surgery is 0.0 ng/ml, and a PSA recurrence is defined as the time after surgery at which the PSA level exceeds 0.2 ng/ml. Genitourinary surgeons and oncologists are particularly interested in whether a surgery using a robotic device improves the time to PSA recurrence compared to non-robotic surgery for removing the cancerous prostate. Because continuous monitoring of PSA is not feasible, the time to PSA recurrence for each patient is interval censored between consecutive blood tests. 
Further, any assumed parametric model for the regression effects of various baseline covariates on risk/hazard of time to PSA recurrence is restrictive. In practice, it is difficult to justify the proportional hazards assumption strictly based on subject-matter knowledge, much less the more restrictive parameteric assumptions which underlie methods such as Weibull regression.
Hence, using a flexible modeling approach which accommodates possible non-linearity and multi-way interactions among the variables is prudent. Moreover, we expect these survival times to be clustered due to the possible presence of unobserved surgeon effects, such as the surgeon's experience, skill, and access/familiarity with modern medical technology. Hence, we need to account for an appropriate within cluster association (in this case, due to surgeon effects) to obtain efficient and appropriate models for survival prediction.
 
 Most semiparametric models for interval-censored survival data use either an Accelerated Failure Time (AFT) model (Hanson et al., 2007) or a proportional hazards model (Sun, 2006; Sinha, Chen and Ghosh, 1999). However, accelerated failure time models do not straightaway give an often-desired interpretation of instantaneous risk and dynamic comparison between two subjects over time. On the other hand, survival data can often violate the restrictive assumption of proportional hazards in practice. In addition, dependence on factors such as experience and skill level of physicians and clinics often lead to clustering of survival data due to random clinician/clinical site effects. Some examples of clustered interval-censored survival data include the National Aeronautics and Space Administration's hypobaric decompression sickness study (HDSD) (Conkin et al., 1992; Conkin and Powell, 2001) and the Lymphatic Filariasis study (Dryer G, Addiss D, 2006). However, failure to account for clustering effects may produce misleading results and, in particular, inaccurate estimation of the precision of the estimated parameters. Most methods for analyzing clustered survival data use frailty random effects (Oakes, 1982) to accommodate within-cluster association as well as between-cluster heterogeneity. Goethals et al. (2009) used a proportional hazards model with a Gamma frailty to analyze clustered interval-censored survival data.

However, main regression effects and effects of interaction among different covariates in survival data are often time-dependent and are potentially more complex than what can be envisioned by a prespecified parametric model of covariate effects on either the hazard, the median, or the mean. This calls for a flexible modeling framework for covariate effects. Some of the proposed approaches to accommodate such complex regression relationships include the boosting proportional hazards model of Li and Luan (2005), bagging (Hothorn et al., 2006) and random survival forests (Ishwaran et al., 2008). Using a Gaussian process to incorporate nonparametric covariate effects, Fernandez et al. (2016) presented a Bayesian semiparametric survival model centered on a parametric baseline hazard model in a fashion similar to ours. 
Their approach makes use of a random feature expansion to approximate the kernel of the Gaussian process for computational purposes; however, the kernel they use for their simulations and data analysis incorporates predictors into the kernel in a restrictive fashion, and essentially restricts the form of their model to a time-varying coefficient model.

Boosting is an example of ensemble learning, which combines multiple weak learners to attain both high stability and a flexible model. The Bayesian additive regression trees (BART) framework proposed by Chipman et al. (2010) is a Bayesian framework for tree-based ensemble models. BART is a computationally efficient and flexible technique, and can accommodate complex non-linear regression relationships. Since its introduction, the BART framework has been extended to account for many different types of model specifications, including semiparametric regression under heteroscedasticity (Pratola et al., 2017), log-linear models (Murray, 2017), multivariate and multi-scale data (Linero et al., 2020), causal inference (Hahn et al., 2020), and varying coefficient models (Deshpande et al., 2020). For systematic reviews of Bayesian tree-based methods and BART, see Linero (2017) and Hill et al. (2020). A recent surge of interest in BART lies in its applications for analyzing survival data. By modeling nonparametric effects of covariates on the survival times through the tree ensembles, Bonato et al. (2011) employed BART for survival prediction in high dimensional genetic studies under three specific models - the proportional hazards regression model, the Accelerated Failure Time (AFT) regression model, and the Weibull regression models where the last one is the special parametric case of former two models. Sparapani et al. (2016) addressed discrete-time survival data with the discrete-time hazard being modeled nonparametrically via BART. However, this approach is aimed essentially at discrete-time survival data and requires the specification of a finite number of possible failure times in order to be used for continuous survival data, and the computational cost increases rapidly for even a moderate increase in the number of possible failure times. Despite this stream of research, to the best of our knowledge, there are no applications of BART for analysis of interval censored survival data. Additionally, two potential shortcomings of the usual BART framework include the lack of smoothness of the estimated regression function and sensitivity to the curse of dimensionality (Linero, 2018). Linero and Yang (2018) recently addressed these drawbacks with an extension of BART called Soft BART (SBART), which provides a considerable improvement over BART by employing an ensemble of ``soft'' decision trees that can adapt to the unknown smoothness level of the true regression function, and can also remove irrelevant predictors.

Other strategies for taking a Bayesian nonparametric approach to this problem have been considered in the literature. For example, the ANOVA-DDP approach of De Iorio et al. (2009) models the survival function using a Dirichlet process mixture whose atoms depend on covariates. The structure of their model effectively assumes that the log survival time density can be expressed as a mixture of normal distributions, with locations of the mixtures being a function of the covariates, but with fixed mixing weights. 

As mentioned by a reviewer, the proposal of this paper is closely related to \emph{hazard regression}, or HARE, approach (Kooperberg et al., 1995). Kooperberg and Clarkson (1997) adapt HARE to address interval censored data, and models the hazard $\log \lambda(t \mid \bf x)$ using a linear spline basis function expansion. A similar approach is taken by Mallick et al. (1999), who use a Bayesian MARS model (which itself is related to the BART models we consider) to perform hazard regression, using log-normal frailties to address clustering. Hazard regression models can be fit in \texttt{R} using the \texttt{polspline} package.
Other tree-based methods exist in the literature for addressing censored survival data with frailties. Su and Tsai (2005) introduce a \emph{tree augmented Cox proportional hazards model}, which uses an auxiliary decision tree to allow for departures from the Cox proportional hazards model. Su et al. (2008) propose the use of \emph{interaction trees} to perform subgroup analysis with censored survival outcomes. Calhoun et al. (2018) implement \emph{multivariate survival trees} (MSTs) to analyze censored survival data, and allow clustering using frailties as we do here. Each of these methods invokes a \emph{single} tree which acts as a strong learner, unlike the BART-based approach we take which uses many decision trees which act as weak learners.

 In this paper, we present a flexible semiparametric model for interval-censored survival data using BART. Our approach models the hazard function of the survival times as a product of (a): a parametric baseline hazard which represents the ``guess/center'' of the actual hazard  and (b): a nonparametric component modeled using SBART to account for the  possible deviation from the guess/center model at (a)  as well as complex time-dependent effects of covariates and cluster-specific random effects on hazard. We use a data augmentation scheme based on ``thinning'' a Poisson process (Adams et al., 2009) to construct an efficient Markov chain Monte Carlo algorithm for sampling from the posterior distribution. The proposed strategy of using a generative model based on sampling from a model defined in terms of rejection sampling can also be used to construct models for conditional distribution estimation; see Li et al. (2020), who construct a BART-based conditional density estimation model using a model defined through rejection sampling from a baseline density. Our proposed model is applicable for left censored, right censored, and interval censored survival data and our computational algorithms are highly scalable to obtain posterior survival prediction. Our simulation studies provide a comprehensive comparison of our proposed model with some existing models for interval-censored survival data. The rest of this paper is organized as follows. In Section~2, we provide a brief review of BART and SBART. In Section~3, we describe our proposed semiparametric model for survival analysis and its extension to censored data, and also provide the data augmentation algorithm used to carry out inference. In Section~4, we present an extensive simulation study to illustrate the performance of the proposed model and compare the results with those obtained from existing methods for survival prediction in interval censored data. In Section~5, we illustrate our methods via analyzing the motivating study of the PSA data. We finally conclude by discussing our key findings in Section~6.

 \section{A brief review of BART and SBART}
 
The BART framework, proposed by Chipman et al. (2010), is a popular Bayesian ensemble method which combines multiple ``weak'' decision trees into a single ``strong'' learner with high predictive accuracy. When the response surface is smooth, as is often the case in practice, BART can be improved theoretically and practically by using the soft BART (SBART) framework introduced by Linero and Yang (2018). SBART makes use of ``soft'' decision trees, and is capable of adapting to the unknown smoothness level of the response surface. We present here a brief review of BART and SBART.
 
\begin{figure}
     \centering
     \includegraphics[width=.8\textwidth]{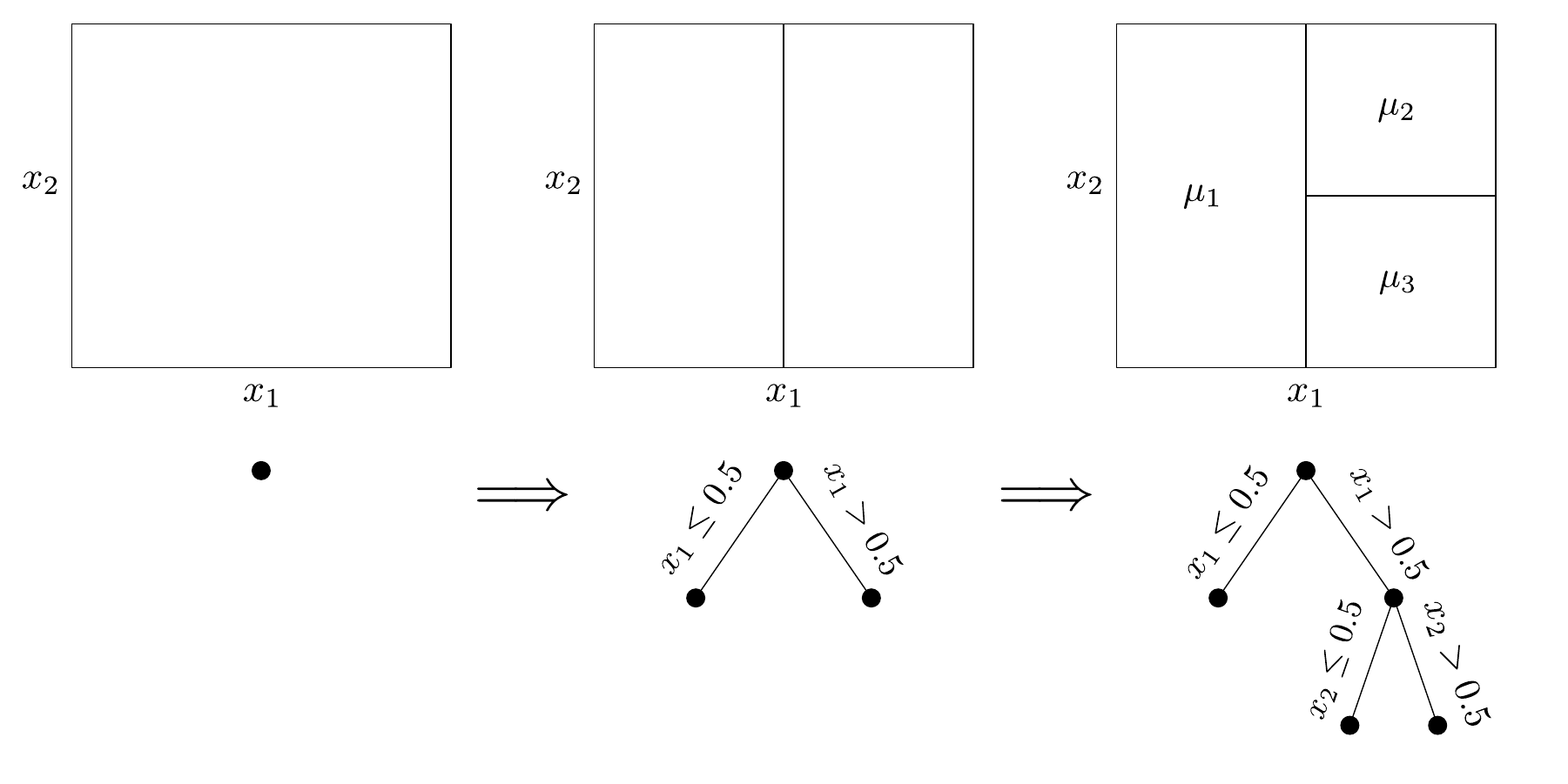}
    \caption{Schematic illustrating how to draw from the prior on $(\tau_t, \mathcal M_t)$. For this tree, we first determine that the root node will be a branch, which occurs with probability $\gamma$; we then sample the splitting coordinate $j = 1$ and the cutpoint $C = 0.5$. This process then iterates; the left child node is set to be a leaf node with probability $1 - \gamma/2^\beta$, and the right child is made a leaf with probability $\gamma/2^\beta$. Eventually this process terminates, and we sample a mean parameter $\mu$ for each leaf node.}
     \label{fig:TreeGrow}
\end{figure}

 Consider a nonparametric regression model $Y = f_0(\mathbf{x}) + \epsilon$ where $\mathbf{x}$ is a $p$-dimensional covariate vector, $f_0:\mathbb{R}^p \to \mathbb{R}$ is an unknown regression function, and $\epsilon \sim N(0,\sigma^2)$ is a random error. BART models the unknown function $f_0(\mathbf{x})$ as a sum of $T$ regression trees given by $f(\mathbf{x}) = \sum_{t=1}^{T}g(\mathbf{x}; \tau_t, \mathcal{M}_t),$ where $\tau_t$ denotes the topology and splitting rules of tree $t$ and $\mathcal{M}_t = (\mu_{t1}, \dots, \mu_{tL_t})$ denotes the set of predictions associated with the $L_t$ terminal nodes of the $t^{\text{th}}$ decision tree. Following Chipman et al. (2010), tree $\tau_t$ is assigned a branching process prior with each node at depth $k=0,1,2,\dots$ being non-terminal with probability $q(k) = \gamma(1+k)^{-\beta},$ where $\gamma > 0$ and $\beta >0$ are positive hyperparameters that control the shape of the tree. For every branch node $b,$ a splitting rule of the form $[x_j \leq C_b]$ is assigned with $\mathbf{x}$ going down left of the tree if the condition is satisfied and right down the tree otherwise. The splitting variable $x_j$ is chosen uniformly from all the $p$ available variables and $C_b$ is assigned a Uniform prior on the set of the available splitting values. Independent Gaussian priors are designated to the terminal node parameters, with $\mu_{tl} \stackrel{iid}{\sim} N(0,\sigma^2_\mu).$ A schematic showing how the branching process prior generates a sample of a decision tree, and its associated partition, is given in Figure~\ref{fig:TreeGrow}.

 While BART is highly flexible and capable of capturing complex regression structures, estimates from BART essentially behave as a sum of step functions, and hence lack smoothness. Note that the function $g(\mathbf x; \tau_t, \mathcal M_t)$ can be written as the step function
 \begin{math}
     g(\mathbf x; \tau_t, \mathcal M_t)
     =
     \sum_{\ell=1}^{L_t} \varphi_\ell(\mathbf x; \tau_t) \, \mu_{t\ell},
 \end{math}
 where $\varphi_\ell(\mathbf x; \tau_t)$ is the indicator that $\mathbf x$ is associated to leaf $\ell$ of $\tau_t$. Using BART to estimate $f(\mathbf x)$ in this fashion would lead to non-smooth estimates, which is not desirable if $f(\mathbf x)$ is believed to be smooth. SBART addresses this drawback by smoothing the weights assigned to each leaf, replacing the indicator function with
 \begin{align*}
     \varphi_\ell(\mathbf x; \tau_t)
     =
     \prod_{b \in \mathcal A_t(\ell)} 
     \psi(x_{j_b} ; C_b, \alpha_b)^{1-R_b} 
     \, 
     \{1-\psi(x_{j_b}; C_b, \alpha_b)\}^{R_b}
 \end{align*}
 where $A_t(\ell)$ denotes the collection of \emph{ancestor nodes} of leaf $\ell$ and $R_b$ is the indicator that the path from the root to $\ell$ at branch $b$ goes right. Here, $\psi(x; c, \alpha)$ is a continuous distribution function of a location-scale family with location $c$ and scale $\alpha$; in this paper, we set $\psi(x; c, \alpha)$ to be the inverse-logit $[1 + \exp\{-(x - c) / \alpha\}]^{-1}$. The parameter $C_b$ plays the role of a cutpoint, while $\alpha$ functions as a bandwidth parameter, with $\alpha \to 0$ corresponding to the usual BART model. When the underlying true $f(\mathbf x)$ is smooth, we expect that SBART will provide lower variance than BART while introducing negligble bias. Additionally, Linero and Yang (2018) show that SBART priors are capable of automatically detecting an appropriate amount of smoothness to use. In the next section, we describe the specifications of our flexible semiparametric model for survival prediction while utilizing this improved prediction performance of SBART to model underlying regression structures.
  SBART also allows for incorporating categorical covariates into the model. As with other BART implementations, SBART binarizes categorical predictors and groups the associated splitting probabilities (with the smoothing of SBART not having a strong consequence for binary variables).

 \section{A semiparametric model for clustered survival data}
 
  Let $T_{ij}$ denote the survival time and $\mathbf x_{ij} = (x_{ij1}, \ldots, x_{ijp}) \in \mathbb R^p$ be a vector of $p$ covariates for subject $j = 1,\ldots,n_i$ in cluster $i = 1,\ldots,N$. Let $f_{ij}$ and $F_{ij}$ be the density and the cumulative distribution function (respectively) of the survival times $T_{ij}$, conditional on the parameters of the model and the random effects. We denote the associated survival function as $S_{ij}(t) = 1 - F_{ij}(t)$ and the hazard function as $\lambda_{ij}(t) = \frac{f_{ij}(t)}{S_{ij}(t)}.$  In this section, we propose a flexible semiparametric model for survival prediction which is well-equipped to capture the complex underlying associations among the different variables in the survival studies, while incorporating unobserved heterogeneity in the population due to random cluster effects. We model the conditional hazard function $\lambda_{ij}(t \mid W_i; \mathbf x_{ij})$ of the survival times $T_{ij}$ given cluster-specific random frailty $W_i$ as 
  \begin{equation}
  	\lambda_{ij}(t\mid W_i;\mathbf{x}_{ij}) = \lambda_0(t) \, W_i \, \Phi(l(t, \mathbf{x}_{ij}))
  	\label{equ:model}
  \end{equation} 
where $\Phi(\cdot)$ is the distribution function of a standard normal random variable and $\lambda_0(\cdot)$ is a parametric baseline hazard on which the model is centered; note, for example, that if $l(t,\mathbf x_{ij}) = 0$, then $\lambda_{ij}(t \mid W_i; \mathbf x_{ij}) = \lambda_0(t) \, W_i / 2$, which is proportional to the baseline hazard $\lambda_0(t)$. The Gamma frailty $W_i \stackrel{ind}{\sim} \Gam(\eta,\eta)$ (Hugaard, 1995) accommodates within-cluster association, and the covariate effects are modeled in a time-varying nonparametric fashion using the SBART prior on $l(t, \mathbf x_{ij})$. 
Note that, due to how SBART is constructed, we have $\sup_{t,\bfx}|l(t,\bfx)|<\infty$ almost-surely so that $\lambda(t \mid W_i; \bfx) \asymp W_i \lambda_0(t)$ almost-surely; this implies that $\int \lambda(t \mid W_i, \bfx)\, dt = \infty$ if-and-only-if $\int \lambda_0(t) \, dt = \infty$, so that our model almost-surely defined a valid hazard function.
Our nonparametric model for $l(t,\mathbf x_{ij})$ allows non-linear effects and interactions among the covariates $\mathbf x_{ij}$ and the survival times $t$. The specification that the shape and rate of the frailty distribution are the same is made to ensure that $E(W_i) = 1$, which is required for identifiability (Hougaard, 1995). The amount of unknown heterogeneity among clusters (surgeons for the PSA study) due to random cluster effects is quantified by the variance $\Var(W_i) = \eta^{-1}$ of the frailties, where a small value of $\eta$ would indicate a large variability among different clusters in the population. As long as $\Lambda_0(t)=\int_0^t\lambda_0(s)ds\to \infty$ as $t\to\infty$, the marginal cumulative hazard function $H_{ij}(t|\mathbf{x}_{ij})=\log S_{ij}(t|\mathbf{x}_{ij})$ resulting from the Gamma frailty model in (\ref{equ:model}) also converges to $\infty$ as $t\to \infty$ because the sample path of $l(s,\mathbf{x}_{ij})$ is bounded with probability $1$. The appeal of our model in equation (\ref{equ:model}) is that it is more semiparametric than usual Cox's proportional hazards model. Because of its semiparametric nature, we do not get the covariate effects as the time-constant hazards ratio of the Cox model that is more parametrically specified. However, we can still nonparametrically estimate the quantities of interest for our clinical collaborators such as the survival curves, median survival times at given values of covariates and time, and restricted mean survival time and even relative risk ratio at a pre-specified time point of interest for given covariate values.

 With interval censored survival data we do not observe $T_{ij},$ but instead, we only record that $T_{ij}$ is in the interval $(A_{ij}, B_{ij}],$ where $A_{ij} \leq B_{ij}$ are the two consecutive inspection times for the unit $(i,j)$. We denote the observed interval censored survival data as  $\mathcal{D} = \{(\mathbf{x}_{ij}, A_{ij}, B_{ij}): j=1, \dots, n_i; i=1, \dots, N\}$ where,  $A_{ij}=B_{ij}$ when  $T_{ij}$ is uncensored, and $B_{ij} = \infty$ when $T_{ij}$ is right censored at $A_{ij}$. Under the hazard function model in (\ref{equ:model}), the likelihood contribution from the unit $(i,j)$, conditional on the unobserved random effect $W_i$, is 
 \begin{align}
 \begin{split}
	\Pr[T_{ij} \in (A_{ij}, B_{ij}] \mid W_i ; \, \mathbf{x}_{ij}] &= S_{ij}(A_{ij}\mid W_i; \, \mathbf{x}_{ij}) - S_{ij}(B_{ij}\mid W_i; \, \mathbf{x}_{ij})\ , 
	\\ \text{where}\ S_{ij}(t\mid W_i; \, \mathbf{x}_{ij}) & 
	=
	\exp\left\{-W_i \int_0^t \lambda_0(s) \, \Phi(l(s,\mathbf{x}_{ij})) \ ds\right\}
	\label{eq:survival}
 \end{split}
\end{align}
is the conditional survival function of $T_{ij}$ given frailty $W_i$. Using the conditional independence of the within-cluster survival times given $W_i$ and across cluster
independence of all survival times, the overall likelihood 
based on all $N$ clusters is then derived as 
 \begin{equation}
  \prod_{i=1}^{N}\int_{0}^{\infty}\left\{\prod_{j=1}^{n_i}	\Pr[T_{ij} \in (A_{ij}, B_{ij}]\mid  W_i;\mathbf{x}_{ij}]
  \right\} g(W_i\mid \eta) \ dW_i 
  \label{equ:like}
\end{equation}
 after integrating out $W_i$ with respect to its $\Gam(\eta,\eta)$ density denoted by  $g(\cdot\mid \eta)$ in (\ref{equ:like}). To fully specify the hierarchical Bayesian formulation of our model, we specify prior distributions of our model parameters as $p(\eta,\Omega, l,\Psi)\propto p(\eta)\times p(\Omega)\times p(l\mid \Psi)\times p(\Psi)$, where $p(\eta)$ is the marginal prior of the shape (as well as the scale) parameter $\eta$ of the Gamma$(\eta,\eta)$ frailty density, $p({\Omega})$ is the prior on the set of unknown parameters ${\Omega}$ associated with the parametric baseline hazard $\lambda_0(\cdot)$, and $p(l|\Psi)$ and $p(\Psi)$ are the priors associated with the SBART model and its hyperparameters. 
 
 In all of our examples, we use a default prior for $\Psi = (\sigma_\mu, \gamma, \beta, r_\alpha)$ which takes $\sigma_\mu = 3 / (k\sqrt m)$,  $(\gamma, \beta) = (0.95, 2)$, and $r_\alpha = 10$ where $\alpha_t \sim \Gam(1,r_\alpha)$ in the SBART prior, with the same bandwidth $\alpha_t$ being used for all branches within a single tree. We use a $\Gam(a,b)$ prior for $\eta$, with $a,b > 0$ carefully chosen to reflect our prior opinion about the extent to which the cluster effects $W_i$ can affect the underlying hazard. By default we set $k = 2$.
 
 We fit this model using Markov chain Monte Carlo; the primary challenge in implementing the model \eqref{equ:like} with the model specified above is computational, as even evaluating \eqref{equ:like} requires numerically evaluating integrals of the form $\int_0^t \lambda_0(s) \, \Phi(l((s, \mathbf x)) \ ds$. As such, it is difficult to construct efficient Markov chain Monte Carlo algorithms for sampling from the posterior distribution by working directly with \eqref{equ:like}. We address this by extending the method of Adams et al. (2009) and Fernandez et al. (2016) to augment the $T_{ij}$'s by viewing them as the first ``accepted'' point of a thinned Poisson process; this will allow us to obtain simple updates for the distributions for all of the model parameters. We elaborate on the implementation of this data augmentation (DA) scheme for uncensored survival times in Section~3.1 and extend it to left-censored, right-censored, and interval-censored survival times in Section~3.2.

\subsection{Data augmentation scheme for no-censoring}

 From the conditional nonparametric hazard function in \eqref{equ:model} and the corresponding survival function \eqref{eq:survival}, it is apparent that the evaluation of the likelihood contribution 
  \begin{equation}
  L_{ij}=W_i \, \lambda_0(T_{ij}) \, \Phi(l(T_{ij},{\mathbf x_{ij}}))\exp\left[
  -\int_0^{T_{ij}}W_i \, \lambda_0(s)\,  \Phi(l(s,{\mathbf x_{ij}})) \, ds\right]
  \label{equ:likij}
 \end{equation}
 for each uncensored $T_{ij}$ case given $W_i$ requires integrating a time-dependent function $\lambda_{ij}(s \mid W_i ; \, \mathbf x_{ij}) = \lambda_0(s) \, W_i \, \Phi(l(s, \mathbf x_{ij}))$; this involves a nonparametrically modeled function $l(s, \mathbf x_{ij})$, which varies by the subjects $(i,j)$. Having this time-varying $l(s, \mathbf x_{ij})$ inside the integral in \eqref{equ:likij} also prevents us from applying the usual Bayesian backfitting algorithm of Chipman et al. (2010) to update $l(s, \mathbf x_{ij})$ given the rest of the parameters. To address these challenges, we use a data augmentation procedure, which introduces additional latent variables $\{G_{ijk} \ \text{for}\ k = 1,\cdots,m_{ij}\}$ to ease the posterior computation (that is, by assuring either closed form or standard updates for the conditional posteriors of the parameters). These random latent variables are generated from a Non-homogeneous Poisson Process (NHPP) with intensity
 \begin{math}
    \lambda_0(s) \, W_i \, \{1 - \Phi(l(s, \mathbf x_{ij}))\}
 \end{math}
 in the interval $(0, T_{ij})$. This process can be sampled by first simulating a Poisson process with intensity function $\lambda_0(s) \, W_i$ on the interval $(0, T_{ij})$ and then ``thinning'' each point with probability $\Phi(l(s, \mathbf x_{ij}))$. Note that the likelihood of the augmented $\mathbf G_{ij}$'s simulated via this NHPP is given by 
 \begin{align}
     \begin{split}
        &\Pr(\text{
           events at $G_{ijk}$ for $k = 1,\ldots, m_{ij}$, and no other events in $(0, T_{ij})$
         })
         \\
         &\qquad=
         \left[\prod_{k = 1}^{m_{ij}} 
         \lambda_0(G_{ijk}) \, W_i \, (1 - \Phi(l(G_{ijk}, \mathbf x_{ij})) \, dG_{ijk}
       \right]
       \times
       e^{-W_i \int_0^{T_{ij}}\lambda_0(s)(1-\Phi(l(s,{\mathbf x_{ij}})))ds}
        \label{eq:augl}
     \end{split}
 \end{align}
  Combining this conditional likelihood given $T_{ij}$ with the likelihood of $T_{ij}$, we get
   \begin{align}
      \lambda_0(T_{ij}) 
        \, W_i 
        \, \Phi(l(T_{ij}, \mathbf{x}_{ij})
      \times \prod_{k=1}^{m_{ij}}
        \lambda_0(g_{ijk})
        \, W_i
        \, \Phi(Z_{ijk} \mid l(G_{ijk}, \mathbf x_{ij}))
      \times 
      e^{-W_i \Lambda_0(T_{ij})},
      \label{equ:augl2}
    \end{align} 
    where $\Lambda_0(t) = \int_0^t \lambda_0(s) \ ds$ denotes the baseline cumulative hazard function of the survival times. This likelihood is greatly simplified because it removes the integral $\int_0^{T_{ij}} \lambda_0(s) \, \Phi(l(s, \mathbf x_{ij})) \ ds$ by adding to it $\int_0^{T_{ij}} \lambda_0(s) \, \{1 - \Phi(l(s, \mathbf x_{ij}))\} \ ds$, cancelling the $\Phi(l(s, \mathbf x_{ij}))$ term. While we have phrased this data augmentation directly in terms of simulating $\mathbf G_{ij}$ from a distribution which simplifies the integral, it is also possible to view this as a traditional data augmentation algorithm by viewing $T_{ij}$ as the first ``accepted'' point from a ``thinned'' Poisson process with intensity $\lambda_0(s) \, W_i$ and acceptance probability $\Phi(l(s, \mathbf x_{ij}))$, with $\mathbf G_{ij}$ being the collection of ``rejected'' points. From this perspective, we are simulating $\mathbf G_{ij}$ from its full conditional distribution, and \eqref{equ:augl2} gives the joint likelihood of the accepted and the rejected points (Fernandez et al., 2016).

The augmented likelihood in \eqref{equ:augl2} is still not ideal because the Bayesian backfitting algorithm of Chipman et al. (2010) requires that the likelihood of $l(t,\mathbf x)$ takes the form of a semiparametric Gaussian regression
\begin{math}
    Z = l(t, \mathbf x) + \epsilon
\end{math}
with $\epsilon \sim N(0, \sigma^2)$; instead, $l(t,\mathbf x)$ enters the  likelihood through the probit terms $\Phi(l(T_{ij}, \mathbf x_{ij}))$ and $1 - \Phi(l(G_{ijk}, \mathbf x_{ij}))$. To accommodate this, we perform another layer of data augmentation using the strategy of Albert and Chib (1993). We introduce truncated normal latent variables $\{Z_{ijk}; k = 1,\ldots, m_{ij}+1\}$ such that
$$
  Z_{ijk} \sim
  \begin{cases}
    N(l(G_{ijk}, \mathbf{x}_{ij}),1)I(-\infty,0), & \text{if}\ k=1,\dots,m_{ij} \\
    N(l(T_{ij}, \mathbf{x}_{ij}),1)I(0,\infty), & \text{if}\ k=m_{ij}+1.
  \end{cases}
$$
After augmenting the $Z_{ijk}$'s to the model, we will use the joint likelihood 
\begin{align*}
    \prod_{i=1}^N 
    \prod_{j=1}^{n_i}
    \left\{
      \lambda_0(T_{ij}) 
        \, W_i 
        \, N(Z_{ij(m_{ij} + 1)} \mid l(T_{ij}, \mathbf{x}_{ij}), 1)
      \times \prod_{k=1}^{m_{ij}}
        \lambda_0(G_{ijk})
        \, W_i
        \, N(Z_{ijk} \mid l(g_{ijk}, \mathbf x_{ij}), 1)
      \times 
      e^{-W_i \Lambda_0(T_{ij})}
     \right\},
\end{align*}
where $N(x \mid \mu, \sigma^2)$ denotes the density of a normal distribution with mean $\mu$ and variance $\sigma^2$. This allows us to utilize the continuous outcome SBART model described by Linero and Yang (2018) by treating the $Z_{ijk}$'s as Gaussian responses with variance $\sigma^2 = 1$. That is, we apply the Bayesian backfitting approach described by Linero and Yang (2018) to the model
$Z_{ijk} = l(G_{ijk}, \mathbf x_{ij}) + \epsilon_{ijk}$ and $\epsilon_{ijk} \sim N(0,1)$,
for $k = 1,\ldots, m_{ij}$, and similarly with $Z_{ijk} = l(T_{ij}, \mathbf x_{ij}) + \epsilon_{ijk}$ for $k = m_{ij} + 1$.

In addition to updating the function $l(t,\mathbf x_{ij})$ using this data augmentation approach, we must also update the frailties $W_i$, the parameter of the frailty density $\eta$, the parameters $\Omega$ of the baseline hazard $\lambda_0(\cdot)$, and the hyperparameters  $\Psi$ of the SBART model. For  the $\Gam(\eta, \eta)$ distributed frailties $W_i$, the likelihood is proportional to
\begin{math}
    W_i^{n_i + \sum_j m_{ij}} \, e^{-W_i \, \sum_j \Lambda_0(T_{ij})}
\end{math}
and the corresponding conjugate conditional posterior is  $\Gam(\eta + n_i + \sum_j m_{ij}, \, \eta + \sum_j \Lambda_0(T_{ij}))$ where $n_i$ is the size of the $i^{th}$ cluster. For the remaining parameters, we use slice sampling (Neal, 2003). 
In summary, our data augmentation scheme is based on the following hierarchical specification of the the model, where each line is conditioned on all of the lines above it and all terms within each line are conditionally independent: 
\begin{align*}
    (\eta, \Psi, \Omega) &\sim p(\eta) \, p(\Psi) \, p(\Omega) \\
    l &\sim \operatorname{SBART}(\Psi) \\
    W_i &\sim \Gam(\eta, \eta) \\
    \text{Data: } T_{ij} &\sim 
      \lambda_{ij}(t \mid \mathbf x_{ij}, l, W_i, \Omega)
      \exp\left\{-\int_0^t \lambda_{ij}(s \mid \mathbf x_{ij}, l, W_i, \Omega) \ ds\right\} \\
    \{G_{ijk}:\ k=1,\cdots,m_{ij}\} &\sim 
      \textnormal{NHPP}\big\{\lambda_0(t) \, W_i \, (1-\Phi(l(t, \mathbf x_{ij})\big\} \qquad \textnormal{for $t \in (0,T_{ij})$} \\
    Z_{ijk} &\sim 
    \begin{cases}
      N(l(G_{ijk}, \mathbf{x}_{ij}),1)I(-\infty,0), & \text{if}\ k=1,\dots, m_{ij} \\
      N(l(T_{ij}, \mathbf{x}_{ij}),1)I(0,\infty), & \text{if}\ k=m_{ij}+1.
    \end{cases}
\end{align*}
The steps of the data augmentation algorithm are summarized in Algorithm~\ref{alg:inference}, which also shows how to sample from the NHPP distribution for $\mathbf G_{ij}$.

\begin{algorithm}[t]
  \caption{Inference algorithm with clustered and uncensored survival times $T_{ij}$.}
  \textbf{Input:} Observed survival times $T_{ij},$ and initiated values of $\Omega$ (baseline hazard parameter), $l$ (SBART), $\Psi$ (parameters of leaf $l$), $W_i$ (frailty variables) and $\eta$ (shape parameter of the frailty density)

  \begin{algorithmic}[1]
  \For{$m = 1, \ldots, S$}
    \For{$i = 1,\ldots,n$}
      \For{$j = 1,\ldots,n_i$}
        \State $q_{ij} \sim \Poisson(1; \Lambda_0(T_{ij} W_i)$
        \State $C_{ij} \sim \Uniform(q_{ij}; \, 0, \,  \Lambda_0(T_{ij})W_i)$
        \State $\tilde{G}_{ij} \gets \Lambda_0^{-1}(\frac{C_{ij}}{W_i})$
        \State $\mathbf{U}_{ij} \sim \Uniform(q_{ij}; \,  0, \, 1)$
        \State $\mathbf{G}_{ij} \gets \{\tilde{G}_{ij}: U_{ij} \le 1-\Phi(l(\tilde{G}_{ij}, \mathbf{x}_{ij}))\}=\{G_{ij1}, \cdots, G_{ijm_{ij}}\}$
        \State $Z_{ijk} \sim \begin{cases}
				N(l(G_{ijk}, \mathbf{x}_{ij}),1)I(-\infty,0), & \text{if}\ k=1,\dots,m_{ij},\\
				N(l(T_{ij}, \mathbf{x}_{ij}),1)I(0,\infty),  & \text{if}\ k=m_{ij}+1.\\
				\end{cases}$
      \EndFor
      \State Update $\Omega$ by slice sampling
	  \State Update $l$ using the Bayesian backfitting algorithm of Linero and Yang (2018)
	  \State Update $\Psi$ by slice sampling
	  \State Update $\eta$ by slice sampling	
	  \State Update $W_i: i = 1, \dots, n$ by sampling $W_i \sim \Gam(\eta + n_i + \sum_j m_{ij}, \eta + \sum_j \Lambda_0(T_{ij}))$
    \EndFor
  \EndFor
  \end{algorithmic}
  \label{alg:inference}
\end{algorithm}

While Algorithm \ref{alg:inference} provides the steps of inference and parameter updates with clustered and uncensored survival times $T_{ij},$ it does not account for the possibility of censoring. In the next sub-section, we discuss the extension of our model for clustered and (left, right or interval) censored survival times.

\subsection{Augmenting $T_{ij}$ for interval censoring}

We now consider the case where $T_{ij}$ is either right or interval censored. Suppose first that $T_{ij}$ is censored at $A_{ij}$, so that we observe the event $[T_{ij} \ge A_{ij}]$. Assuming that the censoring mechanism is non-informative (Kalbfleisch and Prentice, 2002) the targeted likelihood contribution of the event $[T_{ij} \ge A_{ij}]$ is 
\begin{math}
  L_{ij}^* = \exp\left[
    -\int_0^{A_{ij}}
    \lambda_0(s) \, W_i \, \Phi(l(s, \mathbf x_{ij})) \ ds.
    \right]
\end{math}
If we augment $\mathbf G_{ij}$ as before from the NHPP with intensity $\lambda_0(s) \, W_i \, \{1 - \Phi(l(s, \mathbf x_{ij}))\}$ then the corresponding augmented data likelihood contribution is not given by (\ref{equ:augl2}), but instead by 
\begin{align}
  \label{eq:full-likelihood-censored}
  \begin{split}
   L_{ij}^*\times &\Pr[ \textrm{ ``thinned" events at } \mathbf{G}_{ij};
          \textrm{ no other "thinned" events in }(0,A_{ij}]\}
       ]
    \\
    &\qquad= 
       \prod_{k = 1}^{m_{ij}} \{
         \lambda_0(G_{ijk}) \, W_i \, (1 - \Phi(l(G_{ijk}, \mathbf x_{ij})) \ dG_{ijk}
       \}
       \times
       e^{-W_i \Lambda_0(A_{ij})}.
  \end{split}
\end{align}
Hence, we can apply the same data augmentation algorithm as before, where the censored observations contribute \eqref{eq:full-likelihood-censored} to the likelihood. To remove the probit terms from the likelihood, we now only simulate $Z_{ijk} \sim N(l(G_{ijk}, \mathbf x_{ij}), 1) I(-\infty, 0)$ and do not simulate a latent variable for $k = m_{ij}+1$. For the right-censored setting, let $Y_{ij} = T_{ij}$ if $T_{ij}$ is observed and $A_{ij}$ otherwise. Then the resulting likelihood after performing both layers of data augmentation is
\begin{align*}
  &\prod_{i=1}^{N}
    \prod_{j=1}^{n_i}
      \left\{
        \lambda_0(Y_{ij}) \, W_i \, N(Z_{ij(m_{ij}+1)} \mid l(Y_{ij}, \mathbf x_{ij}), 1)
      \right\}^{I(Y_{ij} = T_{ij})}
      \\\qquad&\times
      \prod_{k=1}^{m_{ij}}
        \lambda_0(G_{ijk}) \, W_i \, N(Z_{ijk} \mid l(G_{ijk}, \mathbf x_{ij}), 1)
      \times e^{-W_i \Lambda_0(Y_{ij})}.
\end{align*}

We address interval censoring between $(A_{ij}, B_{ij}]$ by augmenting the true survival time $T_{ij}$; this also addresses left-censoring by setting $A_{ij} = 0$. After augmenting survival time $T_{ij}$, we augment $\mathbf G_{ij}$ using the same approach as for when $T_{ij}$ is observed. The augmented uncensored survival time $T_{ij}$ can be generated as the smallest event time of a non-homogeneous Poisson process with intensity $\lambda_0(t) \, W_i \, \Phi(l(t, \mathbf x_{ij}))$ on the interval $(A_{ij}, B_{ij}]$ \emph{conditional on there being at least one event}. An algorithm for augmenting $T_{ij}$ is given by Algorithm~\ref{alg:imputation}. The time $T_{ij}$ is sampled by first running a Poisson process with intensity $\lambda_0(t) \, W_i$ on $(A_{ij}, B_{ij}]$ and then accepting each point $u$ with probability $\Phi(l(u, \mathbf x_{ij}))$; this process is then repeated until we have at least one acceptance. To make the algorithm more efficient, we condition the initial draw from $\lambda_0(t) \, W_i$ on $(A_{ij}, B_{ij}]$ to have at least one point. It is important to note here, that the imputation step, as described in Algorithm \ref{alg:imputation}, has to be repeated at the beginning of each iteration, for each subject in the dataset, with left-censored or interval censored survival times.

The algorithms discussed in this section are computationally fast. For example, in our simulation study in section \ref{s:sbart-simulation} with  clustered interval censored survival data (under setting D) with 10 clusters and fixed cluster size 10, the algorithms took 0.16 seconds for each iteration using unoptimized \texttt{R} code to perform the data augmentation. Similarly, for the PSA data analysis example in Section \ref{s:sbart-psa_data_analysis} with  597 total observations and 9 clusters,  the algorithm took only 0.22 seconds for each iteration (all computations were performed on a laptop with Intel(R) core i7 processor and 8GB RAM).

\begin{algorithm}[t]
  \caption{Imputation of the survival time for subject $j$ in cluster $i$ with interval censored observations $(A_{ij}, B_{ij}]$}
  \begin{algorithmic}[1]
    \State $m_{ij}^* \gets 0$
    \While{$m_{ij}^* = 0$}
      \State $q_{ij}^* \sim \Poisson(1; W_i(\Lambda_0(B_{ij}) - \Lambda_0(A_{ij})))I(q_{ij}^* > 0)$
      \State $\mathbf{C}_{ij}^* \sim \Uniform(q_{ij}^*; \Lambda_0(A_{ij})W_i, \Lambda_0(B_{ij})W_i)$
      \State $\mathbf{\tilde{G}}_{ij} = \Lambda_0^{-1}(\frac{\mathbf{C}_{ij}^*}{W_i})$
      \State $\mathbf{U}_{ij}^* \sim \Uniform(q_{ij}^*; 0,1)$
      \State $\mathbf{G}_{ij}^* = \{\mathbf{\tilde{G}}_{ij}: U_{ij}^* \le \Phi(l(\mathbf{\tilde{G}}_{ij}, \mathbf{x}_{ij}))\}$
    \EndWhile
    \State Impute $T_{ij}^*$ as $t_{ij}^* = \min\{v : v \in \mathbf G_{ij}^*\}$
  \end{algorithmic}
  \label{alg:imputation}
\end{algorithm}

\subsection{Computational Considerations}

In addition to giving very strong performance across a variety of different problems, methods based on BART are usually more computationally efficient than competing Bayesian nonparametric methods. Several extremely fast implementations of BART exist, including the \texttt{dbarts} and \texttt{bartMachine} packages available on \texttt{CRAN}. A single iteration of the Bayesian backfitting algorithm takes $O(NT)$ computations to compute a Metropolis-Hastings ratio (each observation is dropped down each tree). This is a substantial improvement computationally over the method of Fernandez et al. (2016), who use a crude/restrictive Gaussian process approximation to avoid an $O(N^3)$ matrix inversion.

Relative to standard BART models, the main computational difficulty of our proposed approach lies in the size of the augmented dataset, as the $O(NT)$ complexity per iteration becomes $O(N_{\text{aug}} T)$ where $N_{\text{aug}}$ is $\sum_i n_i + \sum_j m_{ij}$ (assuming no censoring). Hence, the more data that we augment, the slower the computations. While this is not pursued here, this gives additional motivation for choosing a possibly covariate-dependent choice of baseline hazard $\lambda_0(t \mid x)$, as if this is a good approximation of the true baseline hazard then we will not need to augment large datasets. 
An additional computational bottleneck for our approach occurs due to our use of soft trees, which do not permit the same computational shortcuts that standard BART models do. In future work, we hope to develop faster SBART algorithms.

\subsection{The Choice of the Baseline Hazard}

To this point, we have considered using only an exponential or Weibull specification for the baseline hazard $\lambda_0(t)$. Subsequent work by the authors (Linero et al., 2021) gives (among other things) an in-depth examination of the role of the baseline hazard, and develops \emph{covariate-dependent} choices of the baseline hazard $\lambda_0(t \mid \bfx)$. A particularly flexible choice of $\lambda_0(t \mid \bfx)$ they study takes the baseline hazard function \emph{itself} to be modeled using BART. This allows the user, for example, to shrink our nonparametric model towards a semiparametric Weibull proportional hazards model, or towards a semiparametric Cox proportional hazards model with a flexibly-modeled baseline hazard..

A poor choice of $\lambda_0(t \mid \bfx)$ has two consequences for our methods. First, if the shape of $\lambda_0(t \mid \bfx)$ does not match the data well, then the model will compensate by having larger thinning probabilities. This has the consequence of introducing more rejected points into the MCMC scheme, causing inefficient inference. Second, a poor choice of $\lambda_0(t \mid \bfx)$ results in less statistically efficient inferences than would be obtained with a good choice.

 \section{Simulation Study}
 \label{s:sbart-simulation}
 In this section, we illustrate the applicability of our proposed semiparametric model for survival prediction in situations when the regression relationships between the survival times and the covariates are non-linear and complex. We compare the performance of our model with some of the existing survival regression models with readily available computational packages under $4$ different simulation settings:
 \begin{enumerate}
     \item Simulation A: Survival times are independent and uncensored.
     \item Simulation B: Survival times are clustered, but uncensored.
     \item Simulation C: Survival times are independent, but interval-censored.
     \item Simulation D: Survival times are both clustered as well as interval-censored (thus mimicking the motivating post-surgery PSA recurrence study). 
 \end{enumerate}
For each of the above simulation settings, we generate $M=20$ replicates of training datasets, each with $n=100$ subjects, with a subject-specific 5-dimensional  covariate vector $\mathbf{x} = (x_1, x_2, x_3, x_4, x_5)$ drawn from a unit hyper-cube. We first describe the simulation model we use to generate independent and uncensored survival times (for Setting A) and then proceed to describe the extension of this data generation scheme to simulate clustered or/and interval-censored survival times (for Settings B, C and D).

 \textit{Data generation scheme:} We use the common test function
 $$f_0(\mathbf{x}) = 10\sin(\pi x_1 x_2) + 20(x_3 - 0.5)^2 + 10x_4 + 5x_5$$   
 that includes non-linearity and interaction effects among five covariates $(x_1, \ldots, x_5)$. The function $f_0$, proposed initially by Friedman (1993), has non-linear dependence on the first three variables $x_1,$ $x_2$ and $x_3$, linear dependence on $x_4$ and $x_5$, and incorporates a non-linear interaction between $x_1$ and $x_2$. For the $i^{th}$ subject under simulation A, independent survival times $T_i$ are generated from a Gamma distribution with shape parameter $\lambda_i = f_0(\mathbf{x}_i)$ and rate parameter $= 6.$ The observed independent survival data under setting A is $\{\mathbf{x}_i, T_i: i=1, \dots, n\}.$ Predicted survival probabilities are evaluated at a grid of $10$ time points, for each of $n^{*} = 100$ independent subjects in a test dataset generated in the same fashion as the training data. Under simulation setting B, each of our $M=20$ replicated training datasets consists of $N=10$ clusters, with fixed cluster size $n_i = 10.$ We simulate a cluster specific random effect $W_i \sim \Uniform(0,0.2)$ and simulate the survival time for the $(i,j)^{th}$ subject as $T_{ij} \sim \Gam(\lambda_{ij}, 6)$, where $\lambda_{ij} = f_0(\mathbf x_{ij}) + W_i$. Under simulation setting B, we observe $\{\mathbf{x}_{ij}, T_{ij}: i=1, \dots, N, j=1, \dots n_i\}.$

 From the independent survival times $T_i$ (generated as in setting A), we obtain interval censored survival times $(A_{i}, B_{i}]$ (for Simulation C) as follows:
 \begin{itemize}
     \item Generate $K_i \sim \Poisson(T_i)$
     \item Set $A_i = I(K_i = 0) + V_i^{\frac{1}{K_i}}\times T_i \times I(K_i >0),$ where $V_i \sim \Uniform(0,1)$ and $I(\cdot)$ is an indicator function.
     \item Set $B_i = T_i + R_i,$ where $R_i \sim \operatorname{Exponential}(1).$
 \end{itemize}
 This ensures that the censoring mechanism is non-informative. Thus, each replicate of the simulated interval censored survival dataset under setting C is $\{\mathbf{x}_i, A_i, B_i: i=1, \dots, n\}.$ Finally, clustered and interval censored survival times $\{(A_{ij}, B_{ij}]; i=1,\cdots N, j=1,\cdots n_i\}$ (for simulation D) are obtained following the same interval censoring mechanism as above, except now, the censoring is introduced after simulating the clustered survival times $T_{ij}$ (as was obtained for simulation setting B).
 
 \textit{Evaluation of model parameters:} The proposed semiparametric model for survival prediction is fitted to the simulated data in each setting, using a constant baseline hazard $\lambda_0(t) = \Omega,$ that is, assuming that the baseline distribution of the survival times is Exponential with rate $\Omega.$ We assign $\Omega$ a conjugate and relatively non-informative Gamma prior with the shape and the rate hyper-parameters estimated from the data. For analyzing clustered survival times (as in simulation settings B and D), the cluster-specific random effects $W_i$ are assumed to follow a Gamma distribution with equal shape and rate parameter $\eta.$ Note that the amount of unobserved heterogeneity in the population due to the random cluster effect is quantified by the variance $\eta^{-1}$ of the frailty $W_i,$ where a smaller value of $\eta$ indicates a larger cluster effect on the hazard function. However, it is reasonable to believe that the multiplicative effect of the frailties on the estimated hazard function will likely be within $25\%.$ This is ensured by assigning $\eta$ a $\Gam(4, 0.01)$ prior, which allows a probability of less than $0.5\%$ for the frailties to influence the hazard function by more than $25\%.$ A brief analysis of the model sensitivity to different values of the hyperparameter $\eta$ has been discussed in section \ref{s:sbart-psa_data_analysis}. The SBART component of our model ($l$) uses a default prior described by Linero and Yang (2018) with $50$ trees; we find that this default performs reasonably for the survival setting.

We compare our proposed SBART-based semiparametric model for 
survival prediction with parametric Accelerated Failure Time (AFT) models, Bayesian Generalized AFT model (Zhou et al, 2017), frequentist proportional hazards model, random survival forests, and a semiparametric Bayesian proportional hazards model. The parametric AFT models have been fitted using Weibull distributions for the survival times. The Bayesian generalized AFT model (Zhou et al, 2017) as implemented by the \texttt{frailtyGAFT} function in the R package \texttt{SpBayesSurv} allows for a non proportional hazards model with arbitrary censoring, and the semiparametric proportional hazards model (Zhou et al, 2020) as implemented by \texttt{SpBayesSurv} package in R uses the transformed Bernstein Polynomial for fitting the baseline hazard functions and independently distributed Gaussian frailties to model the within cluster association for the clustered survival data (for simulations B and D). Another related model which we could have considered for comparison is the proportional hazards model of Henschel et al. (2009), which can be fit in the \texttt{spatsurv}, \texttt{spatsurv}, and \texttt{R2BayesX} packages.

All Bayesian models were fitted using $2500$ burn-in and $2500$ sampling iterations. Monte Carlo approximation of the root mean squared error of survival prediction of the $j=1,2,\cdots,M=20$ replicate is obtained as 
 $$
 \hat{RMSE}_j 
 = \sqrt{\frac{\sum_{i=1}^{n^*}\sum_{g=1}^{10}\{S(t_g\mid \mathbf{x}_i)-\hat{S}(t_g\mid \mathbf{x}_i)\}^2}{n^*\times 10}},
 $$ 
 where $S(t_g\mid \mathbf{x}_i)$ and $\hat{S}(t_g\mid \mathbf{x}_i)$ are respectively the true and the predicted survival probabilities at time point $t_g$ for the $i^{th}$ subject of the test-dataset, having covariate vector $\mathbf{x}_i.$ For the Bayesian procedures, $\hat{S}(t_g\mid \mathbf{x}_i)$ is taken to be the posterior mean of $S(t_g\mid \mathbf{x}_i).$ Performances of the different methods are compared on the basis of the average root mean squared error, obtained from the 20 replicated simulations and the results are reported in Table \ref{t:tableone}.
 
 As seen from Table \ref{t:tableone}, among the methods considered, our proposed model performs the best in predicting survival functions in terms of the average root mean squared error. It has also been shown that our model is easily implementable with clustered and/or interval-censored survival data. It is worth mentioning here that we found a number of computational packages in R, (for example, \texttt{parfm}, \texttt{frailtypack}) which, although have been built for survival regression models, give frequent issues with parameter convergence and likelihood optimization, especially when the regression relationships among the survival times and the other covariates are complex and non-linear. 
 \renewcommand{\arraystretch}{1.5}
 \begin{table}[t]
	\centering
	\def\~{\hphantom{0}}
		\vspace{0.2in}
		\resizebox{\textwidth}{!}{\begin{tabular}{ccccc }
				\hline 
				
				\multirow{3}{*}{\textbf{Methods}} & \textbf{Simulation A}& \textbf{Simulation B}& 
				\textbf{Simulation C} &
				\textbf{Simulation D}
				\\ 
				
				&(independent, &(independent, &(clustered &(clustered \\
				&uncensored)&interval-censored)&uncensored)&interval-censored)\\
				\hline 
			Parametric AFT&0.7785&0.8793&0.4366&-- \\
			Bayesian generalized AFT &0.1343&0.1198&0.1343&0.1054\\
			Random survival forest &0.1798&--&--&--\\
			Frequentist PH &0.1621&--&0.1313&-- \\
			Semiparametric survival model with SBART &0.1038&0.1106&0.1063&0.0944\\
			Semiparametric Bayesian PH &0.1403&0.1229&0.1418&0.1211\\
			\hline
		\end{tabular}}
    		\caption{\textit{Simulation results based on $20$ replicates of data comparing Monte Carlo estimates of average Root Mean Squared Error (RMSE) obtained from fitting the parametric accelerated failure time (AFT) model with a Weibull distribution for the log survival times, a random survival forest, a Bayesian Generalized AFT model (Zhou et al, 2017), frquentist proportional hazards (PH) model, the proposed semiparametric survival model based on SBART while assuming Gamma frailty and a constant baseline hazard function and a semiparametric Bayesian PH model with Gaussian frailties, and transformed Bernstein polynomials for the baseline hazard.} \label{t:tableone}}
	\vspace*{-6pt}
\end{table}

\section{Application: Prostate Surgery Study}
\label{s:sbart-psa_data_analysis}

A practical example of interval-censored 
clustered survival data is any medical follow-up study to 
to compare the survival functions of the time to prostate specification antigen (PSA) recurrence for competing surgery techniques to remove prostate (Barbash and Glied, 2010).
PSA level in the blood, immediately following surgery, is 0.0 ng/ml; time to PSA recurrence is defined as the time after surgery when the PSA level in the blood exceeds 0.2 ng/ml. Since continuous monitoring of PSA is not feasible, and the level of PSA in the blood can only be determined via blood tests (typically every 3 to 6 months after surgery), the exact time to PSA recurrence is often interval-censored between consecutive post-surgery visits to the clinic. 
Also, these survival times are typically clustered within each 
surgeon/clinic.
The main analysis goal of this study is  
to compare surgery technique groups as well as possible baseline characteristics such as (i) age at surgery (continuous); (ii) number of positive cores (continuous, range 0-17); (iii) Gleason score regarding how the cancer cells are arranged in the prostate (less aggressive or more aggressive cancers); (iv) surgical margin (negative or positive); and (v) pT stage (not spread or spread). 
However, 
due to some errors in the coding of the some of the confounding variables and due to the confidentiality issues, the results of the analysis of the original study cannot be 
presented at this point.  To circumvent this problem, for illustrative purposes and to allow replication of the results using our method, we have created a simulated dataset designed to replicate many of the salient features of the original study. This simulated dataset preserves the number of patients $(n=597)$, number 
and average sizes of clusters (approximately 9 and 66), and number and types of covariates, as well as the performance of our approach relative to competing methods in terms of LPML (our procedure performs better on the real study relative to other methods than it does on the simulated data).  However,  because it is a simulated dataset and the original dataset had coding errors in it, we must emphasize that the analysis results 
of this paper cannot be interpreted as either the observed or estimated effects of the types of surgery. The simulated data does not preserve the effect sizes, and we have blinded all of the confounding and treatment variables with the labels $X_1, \ldots, X_6$.

\begin{figure}[t]
	\centering
	\vspace{0.1in}
    \includegraphics[height = 0.4\textheight]{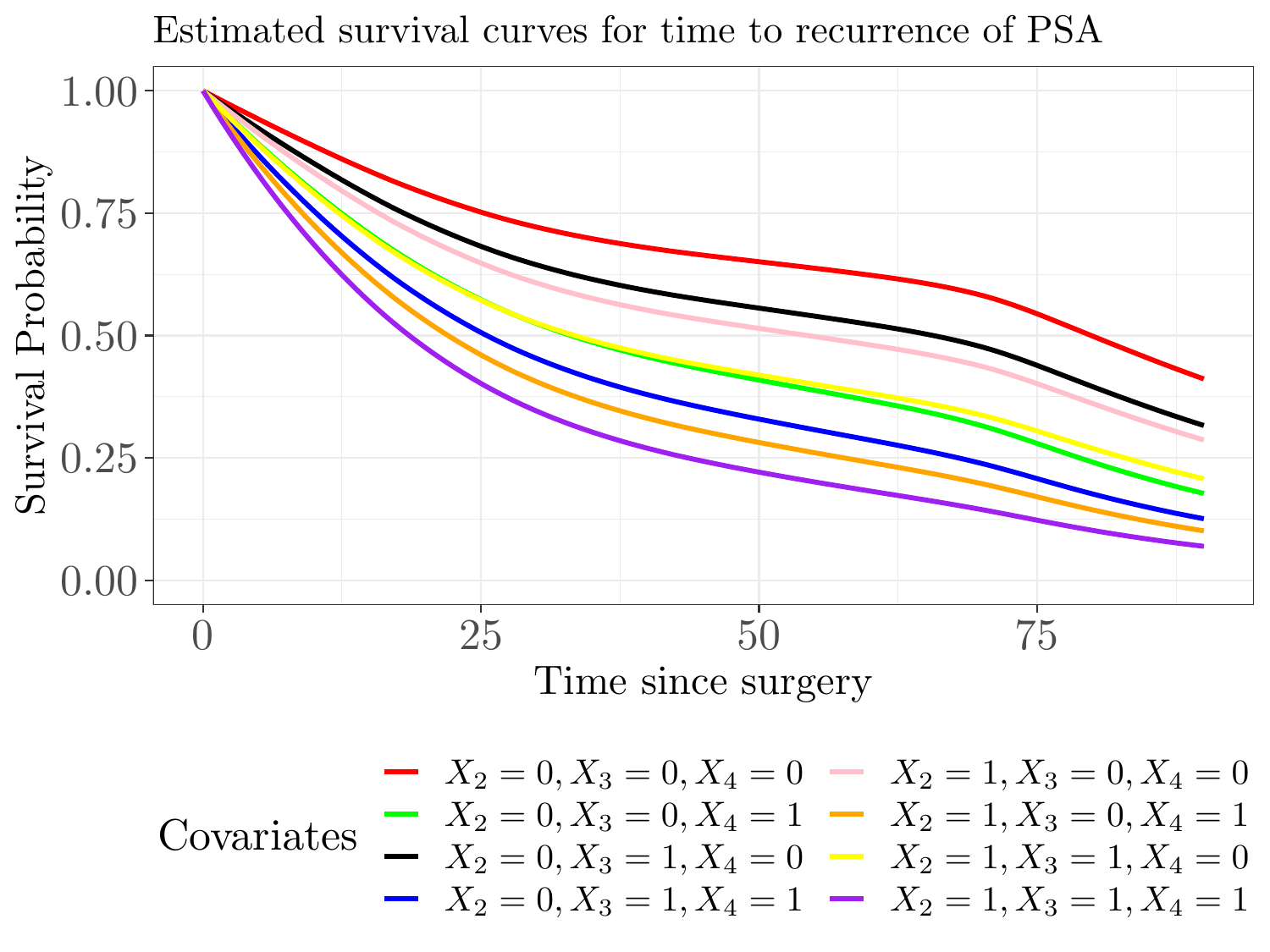}
	\caption{\textit{Estimated survival curves for time to PSA recurrence for a subject who had ``more aggressive" prostate cancer (Gleason score = $1$) with $3$ positive cores and had undergone a prostate removal surgery at $64$ years of age. Covariates shown in the legend correspond to one of two surgical techniques, Surgical margin (positive or negative) and pT stage (Spread or Not spread).} \label{fig:smoothing}}
\end{figure}

Figure \ref{fig:smoothing} shows the estimated survival curves of times to recurrence of PSA for patients with varying values of the covariates $X_2, \ldots, X_4$ (corresponding, in some order, to surgery technique, surgical margin, and pT stage) with $X_1 \equiv 1$  and the two continuous covariates $X_5$ and $X_6$ fixed at their median values. We see that the highest survival times are obtained when each of the is set to $0$, with reasonable consistent drops in survival probability at each time as we change $X_j$ from $0$ to $1$. 

\begin{figure}[t]
	\centering
	\vspace{0.1in}
    \includegraphics[width = 1\textwidth]{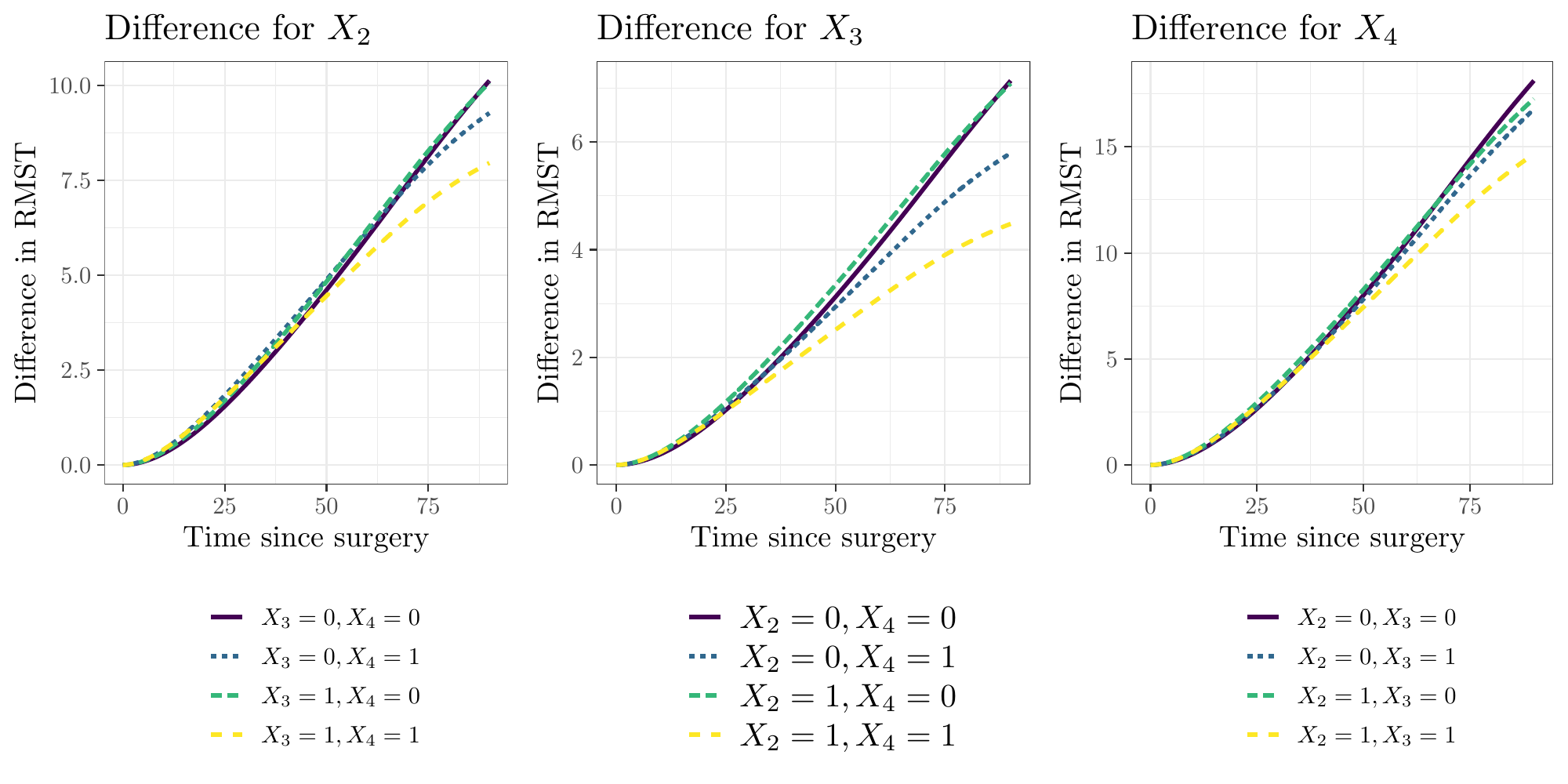}
	\caption{\textit{Difference in Restricted mean survival time (RMST) (measured in months) for time to PSA recurrence due to surgery techniques (technique 1 - technique 0), surgical margin (negative - positive), and pT stage (not spread - spread) (from left to right) for a patient who had a ``more aggressive" prostate cancer (gleason score = 1) with $3$ positive cores and had undergone a prostate removal surgery at $64$ years of age.} \label{fig:rmst}}	
\end{figure}

To assess more fully the effect of each covariate and evaluate potential interactions, we analyze the effect of surgery technique and other covariates on the  time to PSA recurrence using the difference in restricted mean survival time caused by varying the covariate values. Restricted mean survival time (RMST) at a pre-specified time point of interest, $\tau$ is defined as $\RMST(\tau) = E[\min(T,\tau)]$ $= \int_{0}^{\tau}S(u) \ du,$ where $T$ is a non-negative random variable representing the time to event of a particular individual and $S(\cdot)$ is the associated survival function. Inference based on RMST has recently gained popularity, especially due to its ability to summarize survival profiles with non-proportional hazards. Figure~\ref{fig:rmst} shows the differences in RMST to PSA recurrence for different combinations of three binary covariates (surgery technique, surgical margin, and pT stage) for a patient who had a ``more aggressive" prostate cancer (Gleason score = 1) with $3$ positive cores and had undergone a prostate removal surgery at $64$ years of age. The variable with the weakest effect on RMST is $X_3$, which is always estimated to be less than 7 months; the other variables have maximal effects ranging from 10 to 18 months. For variables $X_2$ and $X_4$, the difference in RMST profiles does not appear to interact strongly with the other variables, whereas for $X_3$ we see that the profile differs across levels of $X_2$ and $X_4$. It should be noted that, for the sake of brevity, here we present the estimated survival functions for only certain specific covariate combinations, however, similar analysis and detailed interpretations can be obtained for all other covariates of interest.


Our SBART based semi-parametric model, when fitted to the PSA data set yields the maximum LPML value ($-461.39$) as compared to Zhou et al's (2017) Generalized AFT model with non-proportional hazards (LPML: $-480.19$), a semi-parametric proportional hazards model (LPML: $-471.81$), a semi-parametric proportional odds model (LPML: $-473.35$) with a transformed Bernstein Polynomial for the baseline hazard function, and a semiparametric AFT model (LPML: $-478.32$). These competing methods were fit using the \texttt{spBayesSurv} package in \texttt{R}. This further establishes the goodness of fit of our proposed model for analyzing clustered interval censored survival data with possible non-proportional hazards. 

For the sake of judging the model sensitivity to different hyperparameters, we carried out the PSA data analysis on the original dataset while assigning prior distributions Gamma(5, 0.01), Gamma(10, 0.01), Gamma(7, 0.05) and Gamma(4, 0.05) to the hyperparameter $\eta.$ Note that all the prior choices considered, allow the multiplicative cluster effects to be large ($75\%$). We observed a maximum change of only $4.9\%$ in the predicted survival probabilities of prostate cancer patients who had undergone prostate removal surgeries, the maximum effect of the hyperparameter choice being observed at around 80 months from the time of surgery. This might be attributable to the fact that the motivating dataset only has a few observations with survival times larger than 80 months from the date of surgery. This leads us to conclude that our SBART based semi-parametric model for clustered, interval censored survival data is robust and will perform consistently for any reasonable choice of the hyperparameter $\eta.$

\section{Concluding remarks}
In this article, we have introduced a robust and flexible semiparametric model for clustered, interval-censored survival data under the BART framework. This was accomplished by modeling the hazard function as a product of a parametric baseline hazard and a nonparametric component that uses SBART to incorporate unknown 
time-dependent effects of the covariates and their interactions on the hazard function. Besides being applicable for clustered as well as interval-censored survival data, simulation results validate that our model also attains excellent accuracy in survival prediction, while also requiring minor changes to the usual Bayesian backfitting algorithm used to fit other BART models. Code for fitting the SBART interval-censored survival model with clustering will be made available on the authors' websites. Applicability of the proposed model has been demonstrated via analysis of the motivating post-surgery PSA recurrence study.

The restrictive aspect of our approach is that our frailty model (\ref{equ:model}) assumes that the cluster effects $W_i$ are time-constant and multiplicative on the hazard (similar to frailty model of  Hougaard et al. (1995)). There is some scope for generalizing our approach to allow the frailties to be incorporated more flexibly; at the most extreme end, we might incorporate the frailty into the SBART ensemble. 
However, this comes with additional computational challenges and  the risk of overfitting. We defer determining the best way to incorporate the frailty to future work.

An interesting aspect of our methodology is its adaptability to high-dimensional survival studies involving a large number of covariates and complex underlying associations. Although this paper focuses only on the use of SBART to nonparametrically model the deviations of the hazard function from the baseline hazard, our approach can also be extended to allow for ultra-high dimensional predictors when the function $l(t, \mathbf x_{ij})$ is sparse using the sparsity-inducing Dirichlet prior introduced by Linero (2018), as well as allowing for the penalization of groups of predictors simultaneously (Du and Linero, 2019) in a similar fashion to the group Lasso. This approach might be particularly useful when the primary concern of the study is variable selection, and might provide a new direction for future research in survival prediction.



\begin{thebibliography}{}
\bibitem{ } Adams, R. P., Murray, I. and MacKay, D. J. C. (2011). Tractable nonparametric Bayesian inference in Poisson processes with Gaussian process intensities \textit{ Proceedings of the 26th International Conference on Machine Learning (ICML)}.

\bibitem{ } Albert, J.H., and Chib S. (1993) Bayesian analysis of binary and polychotomous response data. \textit{Journal of the American Statistical Association}, \textbf{88(422)}, 669--679.

\bibitem{} Barbash, G.~I., and Glied, S. A. (2010)
New Technology and Health Care Costs — The Case of Robot-Assisted Surgery, \textit{New England Journal of Medicine},
\textbf{363: 8},
{701-704}.

\bibitem{ } Bonato, V., Baladandayuthapani, V., Broom, M. B., Sulman, E. P., Aldape, K. D. and Do, K., A. (2011). Bayesian ensemble methods for survival prediction in gene expression data \textit{Bioinformatics} \textbf{27:3} 359-–367.


\bibitem{ } Calhoun, P., Su, X., Nunn, M., and Fan, J. (2018). Constructing multivariate survival trees: The MST package for R. \textit{Journal of Statitical Software}, \textbf{83}, 1--21.

\bibitem{ } Chipman, H. A., George, E. I. and McCulloch, R. E. (2002). Bayesian treed models \textit{Machine Learning} \textbf{48,} 299--320.

\bibitem{ } Chipman, H. A., George, E. I. and McCulloch, R. E. (2010). BART: Bayesian additive regression trees \textit{The Annals of Applied Statistics} \textbf{4,} 266--298.

\bibitem{ }Conkin, J., Bedahl, SR and van Liew HD.(1992) A computerized data bank of decompression sickness incidence in altitude chambers.  \textit{Aviation, Space and Environmental Medicine} \textbf{63,} 819--824.
	
\bibitem{ }  Conkin, J. and Powell, M. (2001) Lower body adynamia as a factor to reduce the risk of hypobaric decompression
	sickness. \textit{Aviation, Space and Environmental Medicine} \textbf{72,} 202--214.

\bibitem{ } De Iorio, M., Johnson, W.O., Muller, P., and Rosner, G.L. (2009). Bayesian nonparametric
nonproportional hazards survival modeling. \textit{Biometrics}, \textbf{65}, 762--771.

\bibitem{ } Deshpande, S.K., Bai, R., Balocchi, C., and Starling, J.E. (2020). VC-BART: Bayesian trees for varying coefficients. \textit{arXiv preprint arXiv:2003.06416}.
	
\bibitem{ } Du, J. and Linero, A.R. (2019). Incorporating Grouping Information into Bayesian Decision Tree Ensembles. In \emph{Proceedings of the 36th International
Conference on Machine Learning (ICML)}, 1686--1695.

	
\bibitem{ }	Dreyer, G., Addiss, D., Williamson, J. and Noroes, J. (2006) Efficacy of co-administered diethylcarbamazine and albendazole against adult Wuchereria bancrofti. \textit{Transactions of the Royal Society of Tropical Medicine and Hygiene} \textbf{100,}; 1118–1125.

\bibitem{ }	Fernandez, T., Rivera, N. and Teh, Y. W. (2016) Gaussian processes for survival analysis. \textit{Proceedings of the 30th International Conference on Neural Information Processing Systems} \textbf{16,}; 5021–-5029.

\bibitem{ } Friedman, J. (1991) Multivariate adaptive regression splines. \textit{Annals of Statistics}, \textbf{19}, 1--141.


\bibitem{ } Goethals, K., Ampe, B., Berkvens, D., Laevens, H., Janssen, P. and Duchateau, L. (2009). Modeling interval-censored, clustered cow udder quarter infection times through the shared gamma frailty model. \textit{Journal of Agricultural, Biological and Environmental Statistics} \textbf{14,} 1--14.	

\bibitem{ } Hahn, P.R., Murray, J.S., and Carvalho, C. M. (2020). Bayesian regression tree models for causal inference: regularization, confounding, and heterogeneous effects. \textit{Bayesian Analysis}. To appear.

\bibitem{ }	Hanson, T. and Yang, M. (2007). Bayesian Semiparametric Proportional Odds Models \textit{Biometrics,} \textbf{63,} 88--95.

\bibitem{ } Henschel, V., Engel, J., Hölzel, D. et al. (2009). A semiparametric Bayesian proportional hazards model for interval censored data with frailty effects.\textit{ BMC Med Res Methodol} \textbf{9,} 9.
	
\bibitem{ } Hill, J., Linero, A.R., and Murray, J.S. (2020) Bayesian additive regression trees: a review and look forward. \textit{Annual Review of Statistics and Its Application}, \textbf{7(1)}, 251--278.
	
	
\bibitem{ }	Hothorn, T., Lausen, B. and Benner, A. (2004). Bagging survival trees \textit{Stat Med,} \textbf{23(1),} 77--91.

\bibitem{ }	Hougaard, P. (1995). Frailty models for survival data. \textit{Lifetime Data Analysis,} \textbf{1(3),}255--273.

\bibitem{ } Ibrahim, R., L’Ecuyer, P., Regnard, N. and Shen, H. (2012) On the modeling
and forecasting of call center arrivals. \textit{Proc. 2012 Winter Simulation
Conf., Berlin}, 256--267

\bibitem{ }	 Ishwaran, H., Kogalur, U. B., Blackstone, E. H. and Lauer, M. S. (2008) Random survival forests. \textit{Ann Appl Stat} \textbf{2(3),} 841--860.

\bibitem{ } Kalbfleisch, J. D and Prentice, R. L. (2002) The Statistical Analysis of Failure Time Data, Second Edition. \textit{John Wiley \& Sons}.

\bibitem{ } Kooperberg, C. and Clarkson, D.B. (1997). Hazard regression with interval censored data.
\textit{Biometrics}, \textbf{53}, 1485--1494.

\bibitem{ }	 Li, H. and Luan, Y. (2005) Boosting proportional hazards models using smoothing splines, with applications to high-dimensional microarray data. \textit{Bioinformatics} \textbf{21(10)} 2403-–2409.	 

\bibitem{ } Li, Y., Linero, A.R. and Murray, J.S. Adaptive Conditional Distribution Estimation with Bayesian Decision Tree Ensembles. \textit{arXiv e-prints}.

\bibitem{ } Linero, A. R. (2017). A review of tree-based Bayesian methods. \textit{Communications for Statistical Applications and Methods}, \textbf{24(6)}, 543--559.

\bibitem{ } Linero, A.R. (2018) Bayesian regression tree ensembles that adapt to smoothness and sparsity. \textit{Journal of the American Statistical Association} \textbf{113(522)}, 626--636.

\bibitem{ } Linero, A.R., Basak, P., Li, Y., and Sinha, D. (2021). Bayesian Survival Tree Ensembles with Submodel Shrinkage. \textit{arXiv e-prints}.

\bibitem{ } Linero, A. R., Sinha, D., and Lipsitz, S. R. (2020). Semiparametric mixed‐scale models using shared Bayesian forests. \textit{Biometrics}, \textbf{76(1)}, 131--144.



\bibitem{ } Linero, A. R. and Yang, Y. (2018) Bayesian regression tree ensembles that adapt to smoothness and sparsity \textit{Journal of the Royal Statistical Society. Series B: Statistical Methodology} \textbf{80:5} 1087--1110.

\bibitem{ } Little, R. J. A., and Rubin, D. B. (1987) Statistical analysis with missing data \textit{New York: John Wiley \& Sons}.

\bibitem{} Mallick, B.K., Denison, D.G.T., and Smith, A.F.M. (1999). Bayesian survival analysis using a MARS model. \textit{Biometrics}, \textbf{55}, 1071--1077.

\bibitem{ } Murray, J. S. (2017) Log-linear Bayesian additive regression trees for categorical and count responses. \textit{arXiv preprint arxiv:1701.01503}.

\bibitem{ } Neal, R. M. (2003). Slice sampling. \textit{The Annals of Statistics}, \textbf{31}:705–767.

\bibitem{ } Oakes, D. R. (1982). A model for association in bivariate survival data. \textit{Journal of the Royal
 Statistical Society, Series B}, \textbf{44}: 414--428.
 
\bibitem{ } Pratola, M., Chipman, H., George, E., and McCulloch, R. (2017). Heteroscedastic BART using multiplicative regression trees. \textit{arXiv preprint arXiv:1709.07542}.

\bibitem{ } Sinha, D,. Chen, M-H, and 
Ghosh, S. (1999). Bayesian Analysis and Model Selection for Interval-Censored Survival Data.” \textit{Biometrics}, \textbf{55}, 585–-590.

\bibitem{ } Sparapani, R., Logan, B. R., McCulloch, R. E. and Laud, P. W.  (2016). Nonparametric survival analysis using Bayesian Additive Regression Trees (BART). \textit{Statistics in medicine}.

\bibitem{} Su, X. and Tsai, C.-L. (2005). Tree-augmented Cox proportional hazards models. \textit{Biostatistics}, \textbf{6}, 486--499.

\bibitem{} Su, X., Zhou, T., Yan, X., and Fan, J. (2008). Interaction trees with censored survival data. \textit{International Journal of Biostatistics}, \textbf{4}, 1--26.

\bibitem{ } Sun, J. (2006). The Statistical Analysis of Interval-Censored
	Failure Time Data. \textit{Springer-Varlag}.

\bibitem{} Umlauf, N., Adler, D., Kneib, T., Lang, S., and Zeileis, A. (2015). Structured additive regression models: An R interface to BayesX. \textit{Journal of Statistical Software}, \textbf{63}, 1--46.

\bibitem{} Zhou, H. and Hanson, T. (2018). A unied framework for fitting Bayesian semiparametric models to arbitrarily censored survival data, including spatially-referenced data.
\textit{Journal of the American Statistical Association}, \textbf{113}, 571--581.

\bibitem{} Zhou, H., Hanson, T., and Zhang, J. (2020). spBayesSurv: Fitting Bayesian spatial survival
models using R. \textit{Journal of Statistical Software}, \textbf{92}, 1--33.

\bibitem{} Zhou, H., Hanson, T., and Zhang, J. (2017). Generalized accelerated failure time spatial frailty model for arbitrarily censored data. \textit{Lifetime Data Analysis}, \textbf{23}, 495--515.
\end{thebibliography}
\end{document}